\documentclass[11pt]{article}
\usepackage{amsmath,amssymb,graphicx}
\usepackage{psfrag,subfigure,tabularx}

\def\leigh{Robert G. Leigh}
\def\uiucaddress{\small Department of Physics, University of Illinois, 1110 W. Green St.,
Urbana IL 61801-3080, U.S.A. }
\def\title{\Large { Real-Time Correlators and Non-Relativistic Holography}}

\parskip=\baselineskip

\usepackage{epsfig}
\newcommand{\myfig}[3]{
\begin{figure}[t]
   \epsfxsize=#2cm \centerline{\epsffile{#1}}
   \caption[\arabic{figure}.]{\footnotesize{#3}}
   \label{fig:#1}
\end{figure}
    }

\newcommand\cc[1]{#1^{^{\kern-6pt \circ}}\kern2pt}
\newcommand\comment[1]{}

\font\mybb=msbm10 at 11pt
\def\bb#1{\hbox{\mybb#1}}

\def\bR {\bb{R}}

\newcommand{\h}{\hspace}

\newcommand{\pa}{\partial}

\newcommand{\beq}{\begin{equation}}
\newcommand{\eeq}{\end{equation}}
\newcommand{\beqn}{\begin{eqnarray}}
\newcommand{\eeqn}{\end{eqnarray}}

\def\dalemb#1#2{{\vbox{\hrule height .#2pt
\hbox{\vrule width.#2pt height#1pt \kern#1pt
\vrule width.#2pt}
\hrule height.#2pt}}}
\def\square{\mathord{\dalemb{6.8}{7}\hbox{\hskip1pt}}}

\begin{document}
\renewcommand\author[1]{#1}

\begin{center}
\title
\end{center}
\vskip 2 cm
\centerline{{\bf
\author{\leigh} and
 \author{Nam Nguyen Hoang}
 }}

\vspace{.5cm}
\centerline{\it \uiucaddress}

\vspace{2cm}

\begin{abstract}
We consider Lorentzian correlation functions in theories with non-relativistic Schr\"odinger symmetry.    We employ the method developed by Skenderis and van Rees in which the contour in complex time defining a given correlation function is associated holographically with the gluing together of Euclidean and Lorentzian patches of spacetimes. This formalism extends appropriately to geometries with Schr\"odinger isometry.
\end{abstract}

\pagebreak




\section{Introduction}

Correlation functions of operators in strongly coupled conformal field theories can often be computed using the
AdS/CFT correspondence. Euclidean correlators have a long history\cite{Witten:1998qj,Freedman:1998tz} while the rich analytic structure of various Lorentzian signature correlators can also be obtained. The earliest proposal for the latter was by Son and Starinets\cite{Son:2002sd}, and there have also been several elaborations of that method (see for example \cite{Herzog:2002pc,Iqbal:2009fd}). Recently, Skenderis and van Rees\cite{Skenderis:2008dg,Skenderis:2008dh} showed how the complex time contour of an arbitrary correlation function can systematically be accounted for by gluing together manifolds of various signatures, carefully matching fields at the interfaces. This method was used to calculate scalar two-point functions in AdS space, and in asymptotically AdS spaces.

The extension of gauge-gravity duality ideas to spacetimes of Galilean isometries  and field theories with non-relativistic invariance \cite{Son:2008ye,Balasubramanian:2008dm} has been of much interest in the recent literature. In particular, it is expected that such systems are of more direct relevance to condensed matter models. Correlation functions have recently been computed using standard holographic methods for scalars \cite{Son:2008ye,Balasubramanian:2008dm,Volovich:2009yh} and for fermions \cite{Akhavan:2009ns}.

In this paper, we reconsider {\it Lorentzian} correlators of non-relativistic systems by directly calculating them using the techniques of Refs. \cite{Skenderis:2008dg,Skenderis:2008dh} in Schr\"odinger geometries.
We consider the time-ordered correlator and the Wightman function, as well as thermal correlators.

\section{The Schr\"odinger Geometry and Scalar Fields}

We consider the $d+3$ dimensional Lorentzian geometry\cite{Son:2008ye}
\beq\label{metric}
ds^2 = L^2\left(- b^2\frac{dt^2}{z^4} + \frac{2dt d\xi + d\vec{x}^2 + dz^2}{z^2}\right)
\eeq
where $z\geq 0$ and $b,L$ are length scales.   This geometry has Schr\"odinger isometry with dynamical exponent equal to two. The Killing vectors are of the form
\beqn
N&=&\pa_\xi\\
D&=&z\pa_z+\vec x\cdot\vec\pa+2t\pa_t\\
H&=&\pa_t\\
C&=&tz\pa_z+t\vec x\cdot\vec\pa+t^2\pa_t-\frac{1}{2}(\vec x^2+z^2)\pa_\xi\\
\vec K&=& -t\vec\pa+\vec x\pa_\xi\\
\vec P&=&\vec \pa
\eeqn
$N$ is central, and $D,H,C$ form an $SL(2,\bR)$ algebra.
\comment{The $SL(2,\bR)$ algebra is $[H,D]=2H$, $[H,C]=D$, $[C,D]=-2C$. The quadratic Casimir is ${\cal C}_2=D^2-2(CH+HC)$.}

Consider a massive complex scalar propagating on the
non-relativistic (Lorentzian) geometry
with action
\beq
S = -\frac{1}{2} \int d^{d+3}x \sqrt{-g} \left( g^{\mu\nu}\pa_\mu \bar\phi\pa_\nu \phi +
m_0^2/L^2 |\phi|^2 \right)\label{action}
\eeq
The usual interpretation is that the dual theory lives on $\bR^{1,d}$ at $z=0$ and is coordinatised by the $(t,\vec{x})$ coordinates--$\xi$ is not geometric in the usual sense. The isometry $N:\xi\mapsto\xi+a$ is central and  thus $N$ is strictly conserved. Each operator of the boundary theory can be taken to have a fixed momentum (`particle number') conjugate to $\xi$. $\xi$ is usually taken compact (with circumference $R$) so that the spectrum of possible momenta is discrete. In this case, the dimensionless ratio $b/R$ is a parameter of the theory.

For example, the graviton mode coupling to the stress energy tensor of the boundary theory has
particle number zero \cite{Herzog:2008wg, Adams:2008wt}. Here, we will consider a complex scalar with definite but arbitrary particle number $n$. As we will see, it is very important that the scalar be complex. First, it carries a charge under $N$ and so we should expect it to be complex. More importantly though, it is dual to an operator in a non-relativistic theory, and in such a theory there is a sort of polarization: a simple example of this occurs in free field theories, in which the elementary field creates a particle (and not anti-particle) state.

Now, in this paper we consider correlators of various types. In this regard, as developed by Skenderis and van Rees\cite{Skenderis:2008dg,Skenderis:2008dh}, we regard the metric (\ref{metric}) as defined formally for complex $t$, and a given correlator is constructed from a particular contour in the complex $t$ plane. Here, we consider two such cases, in which the contour is constructed from horizontal (Lorentzian time) and vertical (Euclidean time) contour segments (see Fig. \ref{fig:contour.eps}).
\myfig{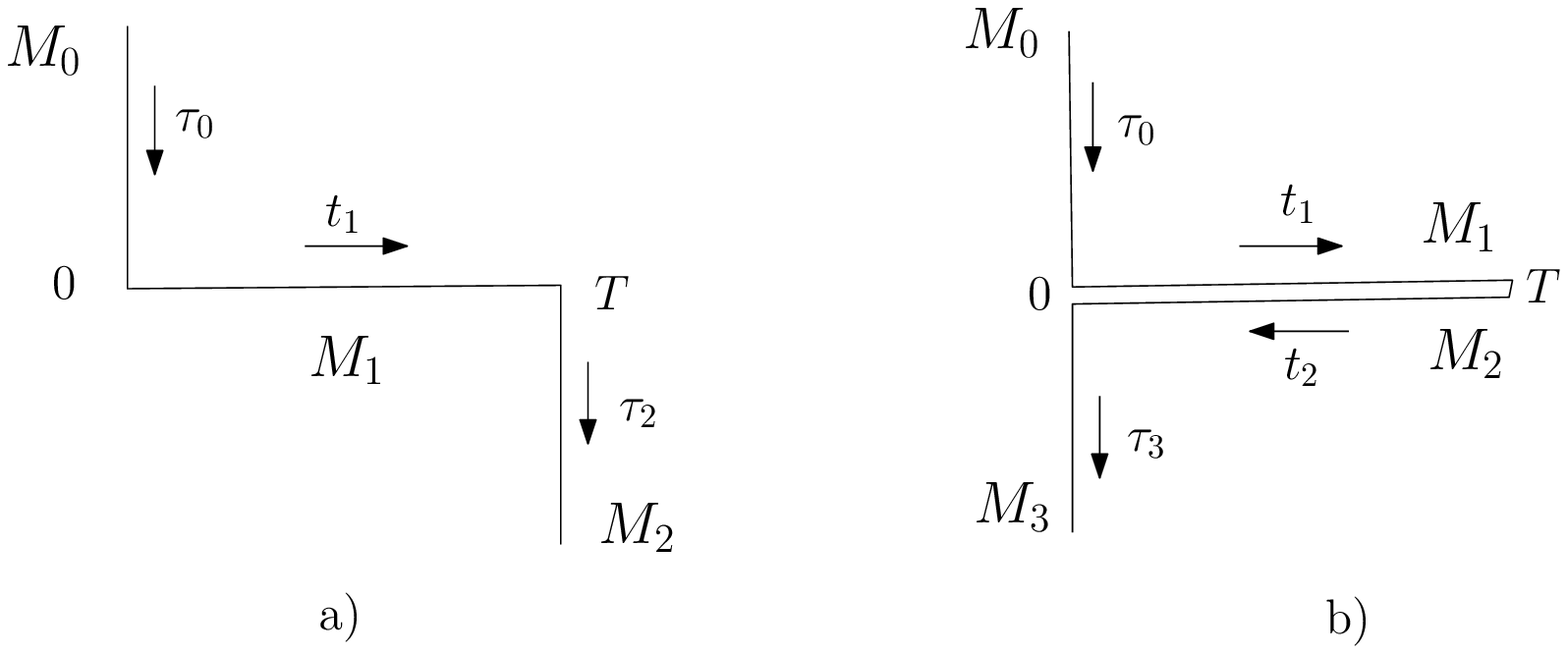}{11}{Contours corresponding to the time-ordered correlator and the Wightman function, respectively.}
%
In the next two subsections, we consider scalar fields in Lorentzian time and in Euclidean time, respectively.

\subsection{Lorentzian signature}

Given the metric (\ref{metric}) for real time, the scalar equation of motion takes the form
\beq
z^2 \pa ^2_z \phi - (d+1)z \pa _z \phi + z^2 (2 \pa_t\pa_\xi + \pa^2_i\phi
) +b^2\pa_\xi^2 \phi - m_0^2 \phi = 0.
\eeq
We look for solutions of the form
\beq
\phi_{(n)} = e^{in\xi} e^{-i\omega t + i \vec k\cdot  \vec x} f_{\omega,n,\vec k}(z),\ \ \ \ \
\bar\phi_{(n)} = e^{-in\xi} e^{i\omega t - i \vec k \cdot \vec x} \bar f_{\omega,n,\vec k}(z)
\eeq
in which case $f$ satisfies
\beq\label{eom}
z^2 \pa^2_z f - (d+1)z \pa_z f + z^2 (2 \omega n - \vec k^2 )f -
m^2 f = 0,
\eeq
where $m^2 = m_0^2 + n^2b^2$. The general solution of (\ref{eom}) can be written in terms of
modified Bessel functions as
\beqn
f_{n,\omega,\vec k}(z) &=& A(\omega,\vec k) z^{\frac{d}{2}+1} K_\nu(q z) + B(\omega,\vec k)
z^{\frac{d}{2}+1} I_\nu(q z)
\eeqn
with $\nu = \sqrt{(\frac{d}{2}+1)^2 + m^2}$ and $q = \sqrt{q^2} =
\sqrt{ \vec k^2-2\omega n}$. $K_\nu$ and $I_\nu$  correspond to non-normalizable and
normalizable modes, respectively. Their asymptotic behavior is as
follows
\beqn
z^{\frac{d}{2}+1} K_\nu(q z \to 0) &=& \Gamma(\nu)
\frac{z^{\frac{d}{2}+1 -\nu}}{2^{-\nu+1}q^\nu} +...\\
z^{\frac{d}{2}+1} I_\nu(q z \to 0) &=& \frac{1}{\Gamma(\nu+1)}
\frac{z^{\frac{d}{2}+1 +\nu}}{2^{\nu}q^{-\nu}} +...\\
z^{\frac{d}{2}+1} K_\nu(|q z| \to \infty) &=& \sqrt{\frac{\pi z^{d+1}}{2q}} e^{-q z}+...\\
z^{\frac{d}{2}+1} I_\nu(|q z| \to \infty) &=& \sqrt{\frac{z^{d+1}}{2\pi
q}} \Big{[}e^{q z}(1+...)+ e^{-q z - i\pi(\nu + 1/2)}(1+...)\Big{]}...
\eeqn
For $q^2 < 0$, both $K_\nu$ and $I_\nu$ are regular everywhere, while
for $q^2 > 0$, $I_\nu$ diverges for large $z$ and should be discarded. This situation is
very similar to that of a scalar field propagating on $AdS_{d+3}$, where the solution can also be
written in terms of modified Bessel functions. In fact this
similarity is very useful and was employed in Ref. \cite{Volovich:2009yh} to
compute the non-relativistic bulk-to-boundary propagator. We note though that there is
a small but important difference due to the non-relativistic nature of
the boundary theory, that we will explain presently.

Without loss of generality, we take $n > 0$. To construct the most general solution (with fixed $n$), we must integrate over all values of $\omega,\vec k$. However, $q$ has a branch point at $\omega=\vec k^2/2n$, and we must then say how to integrate over $\omega$.
Following \cite{Skenderis:2008dg}, we do so by moving the branch point off of the real $\omega$ axis by defining $q_\epsilon = \sqrt{-2\omega n + \vec k^2 - i\epsilon}$, $\bar q_\epsilon =
\sqrt{-2\omega n + \vec k^2 + i\epsilon}$. The branch cut is taken along the negative real axis.
Clearly, we have made a choice here, but we will see later that this is the correct choice, for physical reasons. Notice that since $Re(q_\epsilon), Re(\bar q_\epsilon) > 0$, $K_\nu$ always decays exponentially as $|qz| \to \infty$. In
contrast, the large $z$ behavior of $I_\nu$ tells us that $q, \bar q$
cannot have a real part. As a result, the $i\epsilon$
insertion should not be applied for the normalizable mode.\footnote{This fact was not clearly spelled out in Ref. \cite{Skenderis:2008dg} in the relativistic analogue, but we will see later that it is an important point.}

With these comments, we arrive at  the general solution to (\ref{eom}) in Lorentzian signature
\beqn\label{general solu}
\phi_{(n)}(t,\vec x) &=& e^{in\xi} \int \frac{d\omega}{2\pi} \frac{d^d k}{(2\pi)^d}\
e^{-i\omega t + i \vec k\cdot  \vec x}  z^{\frac{d}{2}+1}
\left( A(\omega,\vec k) K_\nu(q_\epsilon z)
+ \theta(-q^2)B(\omega,\vec k) J_\nu(|q| z) \right)
\eeqn
where we have used $I_\nu(\sqrt{q^2} z) = I_\nu(-i|q|z) \sim J_\nu(|q|z)$.

\subsection{Euclidean signature}\label{sec:eucl}

Next, we consider a similar analysis in Euclidean signature. To do so, we Wick rotate the metric (\ref{metric}) to\cite{Fuertes:2009ex}
\beq\label{E-action}
ds^2 = L^2\left( b^2\frac{d\tau^2}{z^4} + \frac{-2i d\tau d\xi + d\vec{x}^2 + dz^2}{z^2}\right)
\eeq
Although this metric is complex and thus not physical, it is possible to trace carefully through the analysis, and this is
what we need to do in any case for Euclidean signature.

The general solution is
\beqn\label{E-general solu}
\phi_{(n)}(\tau,\vec x) &=& e^{in\xi} \int  \frac{d\omega_E}{2\pi} \frac{d^d k}{(2\pi)^d}\ e^{-i\omega_E \tau +
i \vec k\cdot  \vec x}  z^{\frac{d}{2}+1} A(\omega_E,\vec k) K_\nu(q_E z)\\
\bar\phi_{(n)}(\tau,\vec x) &=& e^{-in\xi} \int  \frac{d\omega_E}{2\pi} \frac{d^d k}{(2\pi)^d}\ e^{i\omega_E \tau
-i \vec k\cdot  \vec x}  z^{\frac{d}{2}+1} \bar A(\omega_E,\vec k)
K_\nu(\bar q_E z)
\eeqn
where now $q_E = \sqrt{q_E^2} = \sqrt{\vec k^2-i2\omega_E n}$. Note that in this case, the branch point is at
imaginary $\omega_E$, and so no $i\epsilon$ insertion is necessary.

In contrast to the Lorentzian case, the Euclidean scalar does not have a
normalizable mode. This is because $q_E$ and $\bar q_E$ cannot be
pure imaginary, so $I_\nu(q_E z)$ is never regular in the interior.
It is important to note, however, that this statement applies to the case $\tau\in (-\infty,\infty)$. If $\tau$ is
restricted, a normalizable mode can emerge. For example, if $\tau \in [0,\infty)$, we write $\omega_E = -i\omega$ for $\phi$ and
$\omega_E = i\omega$ for $\bar\phi$ and the following mode is
allowable
\beqn
\phi &\sim& e^{in\xi} e^{-\omega \tau + i \vec k\cdot \vec x}
z^{\frac{d}{2}+1} I_\nu(q z)\\
\bar\phi &\sim& e^{-in\xi}e^{-\omega \tau - i \vec k\cdot \vec x}
z^{\frac{d}{2}+1} I_\nu(\bar q z)
\eeqn
as long as $\omega > 0$ and $-2\omega n + \vec k^2 < 0$, or
equivalently $\omega > \vec k^2/2n$.

A similar result pertains in the finite temperature case where $\tau\in [0,\beta]$.
Observe however that in contrast to the relativistic real-time formalism, there is no normalizable mode for the Euclidean segment if we restrict $\tau \in (-\infty,0)$. This is because we would need both $\omega < 0$ and
$-2\omega n + \vec k^2 < 0$, and these contradict each other. This will have important consequences. In particular we note that there is no normalizable mode in the segment $M_0$ of either contour in Fig. \ref{fig:contour.eps}.

\section{Non-Relativistic Holography and Correlators}

\subsection{Matching Conditions}

To construct correlation functions, we must match solutions at the interfaces between contour segments. We will label field values on a contour segment $M_n$ by a subscript, $\phi_n$.
Let us begin by considering the Lorentzian$(M_1)$-Lorentzian$(M_2)$ interface in Fig. \ref{fig:contour.eps}b,
where $t_1 \in [0,T]$ and $t_2 \in [T,2T]$ (where $T\to\infty$ is a large time). The total action (for these two segments) is
\begin{align}
S=S_{M_1}+S_{M_2} = \int_0^T dt_1 \left(g^{\mu\nu}_{M_1}\pa_\mu\bar\phi_1 \pa_\nu\phi_1+m_0^2/L^2\bar\phi_1\phi_1\right) - \int_T^{2T} dt_2
\left(g^{\mu\nu}_{M_1}\pa_\mu\bar\phi_2\pa_\nu\phi_2+m_0^2/L^2\bar\phi_2\phi_2\right)
\end{align}
The relative minus sign arises because $M_1$ and $M_2$ have opposite orientation.
For  the same reason, the metric in $M_2$ is
\begin{align}
ds^2_{M_2} = L^2 \Big{(}- \frac{dt_2^2}{z^4} + \frac{-2dt_2 d\xi +
d\vec{x}^2 + dz^2}{z^2}\Big{)},
\end{align}
which has an extra minus sign in the off-diagonal component.

Requiring continuity of the momentum conjugate to $\bar\phi$ 
at the intersection $t_1 = t_2 = T$, we get
\begin{align}
\pa_\xi\phi_1 = \pa_\xi\phi_2.
\end{align}
Along with the continuity of $\phi$, we conclude that the matching
conditions at $t_1 = t_2 = T$ are
\begin{align}\label{matchingcondphi}
\phi_1(T) &= \phi_2(T)\\
n_1 &= n_2\label{matchingcondn}
\end{align}
Thus, we do not need to impose first-order time derivative
continuity of fields along the contour as in the relativistic case --- it is just replaced by particle
number conservation. It turns out that (\ref{matchingcondphi},\ref{matchingcondn}) are also
the matching conditions for Euclidean -- Lorentzian interfaces.

\subsection{Convergence and the Choice of Vacuum}\label{sec:choice}

The non-relativistic holographic correspondence is in general the
same as its relativistic counterpart, where the path integral with
specified boundary conditions in the bulk is identified with the
partition function with sources inserted in the boundary theory. In
the case of a complex bulk scalar, we must temporarily treat the
sources $\phi_{(0)}$ and $\bar\phi_{(0)}$ as independent. The near
boundary expansion of the fields are qualitatively the same as
scalars on $AdS_{d+3}$ \beqn
\phi_{(n)} &=&  e^{in\xi}\Big{\{}\{ z^{\Delta_-} \left(\phi_{(0)}+ z^2 \phi_{(2)} + o(z^4)\right) + z^{\Delta_+} \left(v_{(0)} + z^2 v_{(2)} + o(z^4)\right)\Big{\}}\\
\bar\phi_{(n)} &=&  e^{in\xi}\Big{\{}\{ z^{\Delta_-}
\left(\bar\phi_{(0)}+ z^2 \bar\phi_{(2)} + o(z^4)\right) +
z^{\Delta_+} \left(\bar v_{(0)} + z^2 \bar v_{(2)} +
o(z^4)\right)\Big{\}}, \eeqn with $\Delta_\pm=1+d/2\pm\nu$ and
\begin{align}
\phi_{(2m)} = \frac{1}{2m (2\Delta_+ - (d + 2) -
2m)}\square_0\phi_{(2m-2)},
\end{align}
where here $\square_0 = 2in\partial_t + \partial^2_i$ is the
non-relativistic Laplacian. As usual the holographic correspondence
implies \beq e^{iS^{bulk}_C[\bar\phi_{(0)},\phi_{(0)}]} = \langle
e^{i\int_C ( \hat {\cal O}^\dag\phi_{(0)} + \bar\phi_{(0)} \hat
{\cal O})}\rangle, \eeq where $C$ denotes the contour. Although we
have a very different geometry, it's easily seen that in each patch
of the contour the bulk (either Euclidean or Lorentzian) on-shell
action
\begin{align}
S_{os} = \frac{1}{2}\int_\epsilon d^{d+1}x d\xi\sqrt{|g|}
\hspace{3pt}\bar\phi \hspace{3pt}g^{zz}\hspace{2pt}\partial_z \phi
\end{align}
is essentially the same as scalars on $AdS_{d+3}$. As a result, the
renormalization procedure proceeds in the same way as
$AdS_{d+3}/CFT_{d+2}$, which was carried out in much details in
\cite{Skenderis:2008wp}. In specific, for Lorentzian signature the
counter terms take the form,
\begin{align}
S_{ct} = \int_{\epsilon} d^{d+1}xd\xi \sqrt{-\gamma}\Big{(}\frac{d+2
-\Delta_+}{2}\bar\phi \phi + \frac{1}{2(\Delta_+ - d -
4)}\bar\phi\square_\gamma\phi +\dots \Big{)},
\end{align}
where $\sqrt{-\gamma} = z^{-(d+2)}$ is the $(d+2)$-dimensional
induced metric determinant and $\square_\gamma = z^2 (2in\partial_t
+ \partial^2_i)$ (we will set $L=1$ from now on). The dots represent
higher derivative terms. For special cases where $\nu$ is an
integer, logarithmic counter terms $\sim \log{\epsilon}$ may appear
\cite{Skenderis:2008wp}. It's important to note that $S_{ct}$
preserves the Galilean subalgebra, since $[\square_\gamma, K_i] =
0$. This is in parallel with relativistic holography where the
Poincare subalgebra is preserved by the counter terms. In any case,
$v_{(0)}$ will determine the v.e.v of the dual operator and its
derivative with respect to the source $\phi_{(0)}$ gives us the
2-point functions.

There is, however, a subtlety of which we must be cognizant. Unlike
relativistic field theories, in non-relativistic field theories an
elementary field $\Psi$ and its Hermitian conjugate $\Psi^\dag$ play
the role of creation and annihilation operators. There is a freedom
to choose which is an annihilator, or equivalently a freedom to pick
the vacuum. Once a convention is chosen, $\Psi$ and $\Psi^\dag$ are
no longer on the same footing. This is also true for any operator
$\hat {\cal O}$, $\hat {\cal O}^\dag$, in which $\hat {\cal O}$ is
constructed only from annihilators. This corresponds to the fact
that there is only a single pole in the complex $\omega$-plane in
the non-relativistic case. Consequently, the time-ordered propagator
will in fact have only a single temporal $\theta$-function present.
We expect to see this coming about in the analysis, but to see this
properly, one has to be careful with the convergence of various
integrals.



\section{Correlation Functions}

In both cases shown in Fig. \ref{fig:contour.eps}, we have an initial vertical contour $M_0$. The correlation functions of
interest are computed by including source(s) on horizontal component(s) of the contour.  We first
show that given such a contour component $M_0$, there is no normalizable mode  (such a mode would be everywhere
subleading in the $z\to 0$ expansion). This implies that any
solution with a specific boundary condition is {\it unique}. Indeed, we argued in Section \ref{sec:eucl} that there is
no non-trivial normalizable solution in $M_0$. So in the cases of interest (no sources on $M_0$), $\phi_0 = 0$
identically. The matching condition between $\phi_0$ and $\phi_1$
then requires that $\phi_1(t_1 = 0, \vec x, z) = 0$. The
most general normalizable solution on $M_1$ is
\beq
\phi^{norm}_1(t_1,\vec x, z) =e^{in\xi} \int  \frac{d\omega}{2\pi} \frac{d^d k}{(2\pi)^d}\ e^{-i\omega t_1  + i
\vec k\cdot \vec x}  z^{\frac{d}{2}+1} \theta(-q^2)B(\omega, \vec k)
J_\nu(|q| z).
\eeq
Multiply by $z^{-\frac{d}{2}} e^{-i n \xi -i\vec k'\cdot \vec x} J_\nu(|q'|z)$ with $q'^2 = - 2\omega'
n + \vec k'^2 < 0$ and integrate over $\vec x$ and $z$. We then find
\beq
0 = \int  \frac{d\omega}{2\pi} \frac{d^d k}{(2\pi)^d}d^dx\ e^{i\vec x\cdot (\vec k - \vec
k')} B(\omega, \vec k) \theta(-q^2)\Big{(}\int_0^\infty dz
\h{3pt}zJ_\nu(|q|z) J_\nu(|q'|z)\Big{)}
\eeq
The $z$-integral is elementary (see Appendix, eq. (\ref{1})) and  this becomes
\beqn
0 &=& \int  \frac{d\omega}{2\pi} \frac{d^d k}{(2\pi)^d} d^d x \h{3pt}e^{i\vec x\cdot (\vec k - \vec
k')}
B(\omega, \vec k) \theta(-q^2) \frac{1}{|q'|} \delta(|q| - |q'|)\\
&=& \frac{1}{n}\int  \frac{d\omega}{2\pi} B(\omega, \vec k') \theta(2\omega n - \vec k'^2) \delta(\omega - \omega')\\
&=& \frac{1}{2\pi n} B(\omega',\vec k')\theta(-{q'}^2).
\eeqn
Thus, if $\phi_1(t,\vec x, z) = 0$ at some time, there is no non-trivial
normalizable mode. This reasoning in fact applies  for all segments of both contours in Fig. \ref{fig:contour.eps}.

\subsection{Bulk-Boundary Propagator and Time-ordered Correlator}

Given the absence of a normalizable mode, any solution with sources that we find for
the two contours in Fig. \ref{fig:contour.eps} is unique.
 In this subsection,  we consider contour Fig. \ref{fig:contour.eps}a, with segments $M_0$ ($\tau_0 \in (-\infty, 0]$), $M_1$ ($t_1 \in [0,T]$), $M_2$ ($\tau_2 \in [0,\infty)$).
We place a single $\delta$-function source at $\vec x =
0, t_1 = \hat t_1$ on $M_1$. From our discussions above, $\phi_1$ must be of the form
\beq\label{phi1}
\phi_{1,(n)}(t_1,\vec x,z) =\frac{2}{\Gamma(\nu)}e^{in\xi}  z^{1+d/2} \int  \frac{d\omega}{2\pi} \frac{d^d k}{(2\pi)^d}\ e^{-i\omega (t_1-\hat t_1) + i \vec k\cdot \vec x}
\left(\frac{q_\epsilon}{2}\right)^\nu K_\nu(q_\epsilon z).
\eeq
as this satisfies $\left. z^{-\Delta_-}\phi_{1,(n)}(t_1,\vec x,z)\right|_{z\to 0} = e^{in\xi}\delta(t_1-\hat t_1)\delta(\vec x)$, and any ambiguity corresponds to normalizable modes, which we have argued are zero.
Since there are no sources on $M_2$, $\phi_2$ takes the form
\beq
\phi_{2,(n)} = \frac{2\pi i}{\Gamma(\nu)} e^{in\xi} z^{1+d/2} \int  \frac{d\omega}{2\pi} \frac{d^d k}{(2\pi)^d}\ e^{-\omega (\tau + iT-i\hat t_1)  + i \vec k\cdot \vec x}
\theta(-q^2)\left(\frac{|q|}{2}\right)^\nu J_\nu(|q|z).
\eeq
which has been deduced from the matching condition $\phi_1(t_1=T) = \phi_2(\tau = 0)$ as follows. For any time $t_1 >\hat t_1$, we can re-expand $\phi_1$ in terms of $J_\nu$'s. In particular, at $t_1= T$,
we should have
 \beq
\int \frac{d\omega}{2\pi} \frac{d^d k}{(2\pi)^d}\ e^{-i\omega (T-\hat t_1)  + i
\vec k\cdot \vec x}  q^\nu_\epsilon zK_\nu(q_\epsilon z)
= \int \frac{d\omega}{2\pi} \frac{d^d k}{(2\pi)^d}\ e^{-i\omega (T-\hat t_1)  + i
\vec k\cdot \vec x} C(\omega, \vec k)
\theta(-q^2)zJ_\nu(|q|z)
\eeq
for some $C(\omega, \vec k)$. To find this coefficient we use the
same trick as in the last subsection: multiply both sides by
$e^{i \omega'(T - \hat t_1) -i\vec k' \vec
x} J_\nu(|q'|z)$ with $q'^2 = - 2\omega' n + \vec k'^2 < 0$ and
integrate over $\vec x, z$. The right-hand side gives $\frac{1}{2\pi n}
\theta(-q^2)C(\omega',\vec k')$, while the left-hand side can be computed using
(\ref{2}) to give $\frac{i}{2n} |q'|^\nu$.

The bulk-boundary propagator
is essentially identified with $\phi_1$ itself: if we simply strip off the $e^{in\xi}$ factor, we can write
\beqn
K_{n,n'}(t,\vec x,z) &=& \delta_{n,n'} K_{(n)}(t,\vec x,z)\\
K_{(n)}(t,\vec x,z;\hat t)&=&\frac{2 z^{1+d/2}}{\Gamma(\nu)}  \int \frac{d\omega}{2\pi} \frac{d^d k}{(2\pi)^d}\ e^{-i\omega (t-\hat t) + i \vec k\cdot \vec x} \left(\frac{q_\epsilon}{2}\right)^\nu
K_\nu(q_\epsilon z).
\eeqn
As shown in Ref. \cite{Volovich:2009yh} for example, this is closely related to the bulk-boundary propagator in $AdS_{d+3}$. Alternatively, we may perform the integration directly, following the analogous treatment
in Ref.  \cite{Skenderis:2008dg}. To do so, it is convenient to convert the $\omega$-integral to an integration over $p=q_\epsilon$, and the contour in the $p$-plane is as shown in Fig. \ref{fig:bulk-boundary-contour.eps}.
\myfig{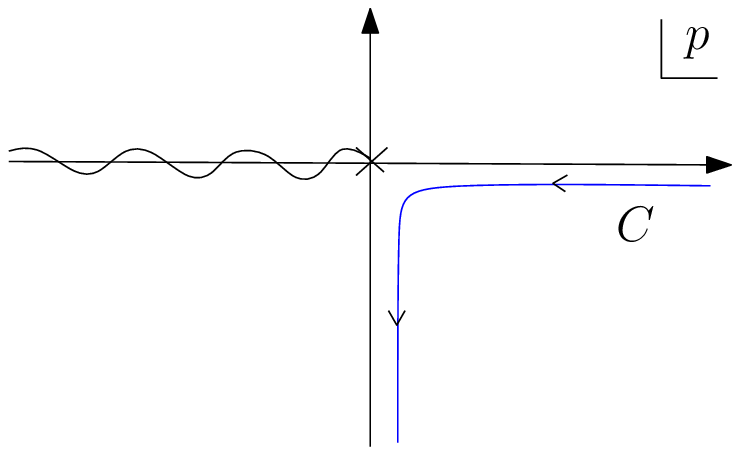}{6}{Contour of integration in the complex $p$-plane for the Lorentzian bulk-boundary propagator.}
Here though there is just one branch point (at $\omega=\vec k^2/2n-i\epsilon$) and the $i\epsilon$ tells us in which sense to traverse the cut. One arrives at
\beq\label{bulk-boundary}
K_{(n)}(t,\vec x,z;\hat t)= \theta(t_1-\hat t_1)\frac{1}{\pi^{d/2}\Gamma(\nu)}\left(\frac{n}{2i}\right)^{\Delta_+-1}
\left(\frac{z}{t_1-\hat t_1}\right)^{\Delta_+} e^{in\frac{z^2 + \vec x^2 +i\epsilon}{2(t_1-\hat t_1) }}
\eeq
where $\Delta_\pm=1+d/2\pm\nu$.

The correlator is then identified with the $z^{\Delta_+}$
coefficient in the near boundary expansion of $\phi_1$ (without the
$e^{in\xi}$ factor) \beq\label{time-ordered} \langle T\Big{(}\hat
{\cal O}_{(n)}(\vec x,t_1) \hat {\cal O}_{(n)}^\dag(\vec
x',t'_1)\Big{)}\rangle =
\frac{1}{\pi^{d/2}\Gamma(\nu)}\left(\frac{n}{2i}\right)^{\Delta_+-1}
\frac{\theta(t_1-t_1')}{(t_1- t_1')^{\Delta_+}} e^{in\frac{(\vec x -
\vec x')^2 + i\epsilon}{2(t_1-t_1')} }. \eeq

\subsection{Wightman function}

The time-ordered correlator, as we have explained, contains a single temporal $\theta$-function. It does not tell us about $\langle
\hat {\cal O}(\vec x,t_1) \hat {\cal O}^\dag(\vec x',t_1')\rangle$ for $t_1' > t_1$. To find this
2-point function we work with the contour of Fig.  \ref{fig:contour.eps}b. Denote the
segments by $M_0$ ($\tau_0 \in (-\infty, 0]$), $M_1$ ($t_1 \in
[0,T]$), $M_2$ ($t_2 \in [T,2T]$) and $M_3$ ($\tau_3 \in
[0,\infty)$) as sketched in the figure. We place a $\delta$-function
source at $\vec x = 0, t_1 = \hat t_1$ on $M_1$ and nowhere else. The Wightman function is obtained then from $\phi_2$, the field on $M_2$.
Here $\phi_0 = 0$ and $\phi_1$ remain the same as (\ref{phi1}).
Given experience from the last subsection, we can see immediately
that $\phi_2$ should be
\begin{align}
\phi_{2,(n)} = \frac{2\pi i}{\Gamma(\nu)} e^{in\xi} z^{1+d/2}\int \frac{d\omega}{2\pi} \frac{d^d k}{(2\pi)^d}\ e^{-i\omega (2T - t_2-
\hat t_1) + i \vec k\cdot \vec x} \left(\frac{|q|}{2}\right)^{\nu}
\theta(-q^2)J_\nu(|q|z).
\end{align}
This has been determined by requiring the matching condition $\phi_1(t_1 = T) = \phi_2(t_2 = T )$. Notice the unusual $e^{+i\omega t_2 + i\vec k\cdot \vec x}$ wave factor. It is related to the fact mentioned before that along this part of the contour, the metric has an extra minus
sign in the $g_{t_2\xi}$ component.

It is now necessary to compute $\phi_2$ in coordinate space.
We make a change of variable $p = |q| = \sqrt{2\omega n- \vec k^2}$
\beq
\phi_2= \frac{i}{n\Gamma(\nu)2^\nu} e^{in\xi} z^{1+d/2}
\int_0^\infty dp\ e^{-ip^2 (2T-t_2-\hat t_1)/2n}  p^{\nu+1} J_\nu(pz)
\int \frac{d^d k}{(2\pi)^d}\ e^{-ik^2 (2T-t_2-\hat t_1)/2n}e^{i \vec k\cdot \vec x}.
\eeq
We note that both integrals converge if $2T-t_2-\hat t_1\to 2T-t_2-\hat t_1-i\epsilon$. The first integral can be computed using (\ref{3}), while the second
one is just a Gaussian integral. The final result is
\beq\label{phi2}
\phi_2 = e^{in\xi}\frac{1}{\pi^{d/2}\Gamma(\nu)}\left(\frac{n}{2i}\right)^{\Delta_+-1}
\Big{(}\frac{z}{\tilde t_2-\hat t_1-i\epsilon}\Big{)}^{\Delta_+}
e^{in\frac{z^2 + \vec x^2}{2(\tilde t_2-\hat t_1 - i\epsilon) }}.
\eeq
where $\tilde t_2=2T-t_2$. Observe that $\phi_2$ is closely related to the bulk-boundary propagator
(\ref{bulk-boundary}) except for the absence of the step function
and a different $i\epsilon$ insertion, as expected.

The vacuum expectation value of $\hat {\cal O}(\tilde t_2,\vec x)$ is
\begin{align}
\langle \hat {\cal O}(\tilde t_2,\vec x) e^{i(\phi_{1(0)} \hat {\cal O}^\dag +
\bar\phi_{1(0)} \hat {\cal O})}\rangle =\frac{1}{\pi^{d/2}\Gamma(\nu)}\left(\frac{n}{2i}\right)^{\Delta_+-1}
 \int d t_1 d^d x'\frac{e^{in\frac{(\vec x-\vec x')^2}{2(\tilde t_2- t_1 - i\epsilon) }}}{(\tilde t_2-t_1-i\epsilon)^{\Delta_+}}
 \phi_{1(0)}(t_1, \vec x') .
\end{align}
Taking a derivative with respect to $\phi_{1(0)}$ and setting the source to zero,
we get the Wightman function
\beq
\langle \hat {\cal O}(\tilde t_2,\vec x) \hat {\cal O}^\dag(t_1, \vec x') \rangle
=\frac{1}{\pi^{d/2}\Gamma(\nu)}\left(\frac{n}{2i}\right)^{\Delta_+-1}
\frac{e^{in\frac{(\vec x-\vec x')^2}{2(\tilde t_2- t_1 - i\epsilon) }}}{(\tilde t_2-t_1-i\epsilon)^{\Delta_+}}
\eeq
Notice that $\hat {\cal O}^\dag$ is always in the front of $\hat {\cal O}$ because
$t_1$ is always the earlier contour time.

\subsection{Thermal Correlator}

Finally, we compute a thermal correlator by taking the time direction to be compact of period $\beta$.
\myfig{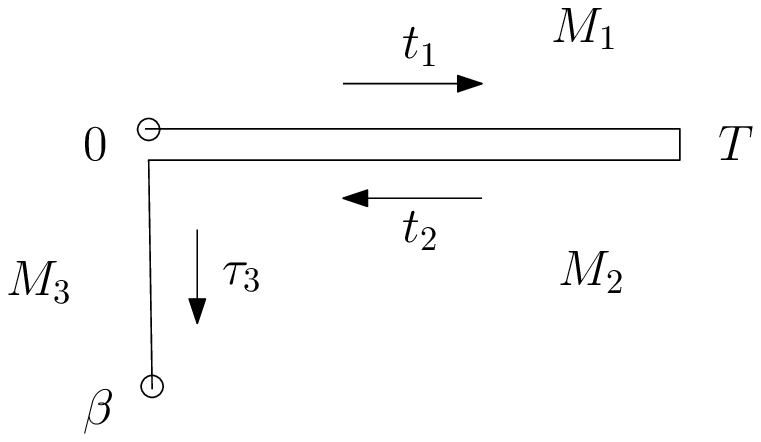}{7.5}{Thermal contour. Points with a circle are identified.}

To compute the thermal time-ordered correlator and Wightman function, we
consider the thermal contour shown in Fig. \ref{fig:thermal-contour.eps}, where $t = 0$ and $t =
-i\beta$ are identified. We place a $\delta$-function
source at $t_1 = \hat t_1, \vec x = 0$. Note that in contrast to the previous discussions, here there is no $M_0$ component of the contour. It is convenient in this context to write the general solution along $M_1$ in the form
\beq\label{phi1T}
\phi_1 = \frac{2 e^{in\xi}z^{1+d/2}}{\Gamma(\nu)}\int \frac{d\omega}{2\pi} \frac{d^d k}{(2\pi)^d}\ e^{-i\omega (t_1 - \hat t_1)+ i\vec k\cdot\vec x }\Big{(} A(\omega,\vec k) \left(\frac{q_\epsilon}{2}\right)^\nu K_\nu(q_\epsilon z) + B(\omega,\vec k)\left(\frac{q_{-\epsilon}}{2}\right)^\nu K_\nu(q_{-\epsilon} z)\Big{)}.
\eeq
where $q_{-\epsilon} = \bar q_\epsilon = \sqrt{-2\omega n + \vec
k^2 + i\epsilon}$.
In order that this correspond to a $\delta$-function source for $z\to 0$, we must have $A + B = 1$. (Furthermore, the case $B=-A$ corresponds to a normalizable mode.) Note that because of the condition on $A,B$, although $A$ and $B$ are not necessarily analytic functions, their sum is analytic. Thus for example, for any pole in $A$, there will be a corresponding pole in $B$ with opposite residue. All of their poles will
contribute opposite residues and cancel out each other in the limit
$\epsilon \to 0$. In (\ref{phi1T}), the first term has support for $t_1>\hat t_1$, while the second has support for $t_1<\hat t_1$.

The matching condition at $(M_1,M_2)$ and $(M_2,M_3)$ intersections imply that
\beqn
\phi_2 &=&\frac{2\pi i e^{in\xi}z^{1+d/2}}{\Gamma(\nu)}
 \int\frac{d\omega}{2\pi} \frac{d^d k}{(2\pi)^d}\  e^{-i\omega (2T - t_2 -\hat t_1) + i\vec k\cdot\vec x }
 A(\omega,\vec k) \left(\frac{|q|}{2}\right)^\nu J_\nu(|q| z)\theta(-q^2)\label{phi2T}\\
\phi_3 &=&\frac{2\pi i e^{in\xi}z^{1+d/2}}{\Gamma(\nu)}
\int \frac{d\omega}{2\pi} \frac{d^d k}{(2\pi)^d}\ e^{-\omega (\tau_3 - i\hat t_1)  + i\vec k\cdot\vec x }
A(\omega,\vec k) \left(\frac{|q|}{2}\right)^\nu J_\nu(|q| z)\theta(-q^2)\label{phi3T}
\eeqn
The thermal condition $\phi_1(t_1 = 0) = \phi_3(\tau_3 = \beta)$
along with $A+B=1$
then gives
\beq
A = \frac{1}{1- e^{-\beta \omega}}, \h{20pt} B = \frac{1}{1 -e^{+\beta \omega}}.
\eeq
As usual, the time-ordered propagator is the coefficient of
$z^{\Delta_+}$ in the small $z$ expansion of $\phi_1$ (without the
$e^{in\xi}$ factor). Hence we get\footnote{For integer $\nu$, there is an extra logarithmic factor, namely
$q_{\pm\epsilon}^{2\nu}$ is replaced by $q^{2\nu}_{\pm\epsilon}
\ln{q_{\pm\epsilon}^2}$.}
\beq\label{time-orderedT}
\langle T\Big{(}\hat O(x) \hat O^\dag(x') \Big{)}\rangle \sim \int
\frac{d\omega}{2\pi} \frac{d^d k}{(2\pi)^d}\ e^{-i\omega (t - t') + i\vec k\cdot(\vec x - \vec x') }
\Big{(} \frac{(-2\omega n + \vec k^2 - i\epsilon)^{\nu}}{1-e^{-\beta \omega}}
+ \frac{(-2\omega n + \vec k^2 +i\epsilon)^{\nu}}{1 - e^{\beta \omega}} \Big{)}.
\eeq
Note that this has the expected form for a thermal correlator\cite{Skenderis:2008dg}
\beq
\langle T\Big{(}\hat O(x) \hat O^\dag(x') \Big{)}\rangle = -N(\omega)\Delta_A(\omega,\vec k)+(1+N(\omega))\Delta_R(\omega,\vec k)
\eeq
In the present notation, $N=-B$. We can also write this as the zero temperature result plus a finite temperature piece:
\beq
\langle T\Big{(}\hat O(x) \hat O^\dag(x') \Big{)}\rangle\sim \int \frac{d\omega}{2\pi} \frac{d^d k}{(2\pi)^d}\ e^{-i\omega (t - t') + i\vec k\cdot(\vec x - \vec x') }\left[ q_\epsilon^{2\nu} - \frac{1}{1-e^{\beta\omega}}(q_\epsilon^{2\nu}-q_{-\epsilon}^{2\nu})\right]
\eeq

The Wightman function can also be read off from $\phi_2$
\begin{align}
\langle \hat {\cal O}(x) \hat {\cal O}^\dag(x') \rangle \sim i\pi \int \frac{d\omega}{2\pi} \frac{d^d k}{(2\pi)^d}\ e^{-i\omega (t -  t' - i\epsilon) + i\vec k(\vec x
-\vec x')} \frac{(2\omega n - \vec k^2)^\nu}{1- e^{-\beta \omega}}
\theta(2\omega n - \vec k^2)
\end{align}

\appendix
\section{Appendix}

We record integrals that have been useful in the above analysis.
\begin{align}\label{1}
\int_0^\infty t\h{2pt}J_\nu(qt) J_\nu(q't)\h{2pt}dt =
\frac{1}{q}\delta(q-q'),\h{30pt} q,q' \text{real}, \h{3pt}\nu >
-\frac{1}{2}
\end{align}
\begin{align}\label{2}
\nonumber &\int_0^\infty K_\mu(at)J_\nu(bt) t^{\mu+\nu +1} dt =
\frac{(2a)^\mu
(2b)^\nu \Gamma(\mu+\nu+1)}{(a^2+b^2)^{\mu+\nu+1}},\\
&\h{150pt} Re(\nu+1)
> Re(\mu), \h{3pt}Re(a) > |Im(b)|
\end{align}
\begin{align}\label{3}
\int_0^\infty e^{-a^2 t^2} t^{\nu+1} J_\nu(bt) dt =
\frac{b^\nu}{(2a^2)^{\nu+1}} e^{-\frac{b^2}{4a^2}},\h{30pt} Re(\nu)
> -1, \h{3pt} Re(a^2) > 0
\end{align}

\providecommand{\href}[2]{#2}\begingroup\raggedright\endgroup

\end{document}